# Fluidic bypass structures for improving the robustness of liquid scanning probes

David P. Taylor[1,2], Govind V. Kaigala[2]*, Senior Member, *IEEE*

*Abstract— Objective*: We aim to improve operational robustness of liquid scanning probes. Two main failure modes to be addressed are an obstruction of the flow path of the processing liquid and a deviation from the desired gap distance between probe and sample. *Methods*: We introduce a multi-functional design element, a microfluidic bypass channel, which can be operated in dc and in ac mode, each preventing one of the two main failure modes. *Results*: In dc mode, the bypass channel is filled with liquid and exhibits resistive behavior, enabling the probe to passively react to an obstruction. In the case of an obstruction of the flow path, the processing liquid is passively diverted through the bypass to prevent its leakage and to limit the build-up of high pressure levels. In ac mode, the bypass is filled with gas and has capacitive characteristics, allowing the gap distance between the probe and the sample to be monitored by observing a phase shift in the motion of two gas-liquid interfaces. For a modulation of the input pressure at 4 Hz, significant changes of the phase shift were observed up to a gap distance of 25 μm. *Conclusion*: The presented passive design element counters both failure modes in a simple and highly compatible manner. Significance: Liquid scanning probes enabling targeted interfacing with biological surfaces are compatible with a wide range of workflows and bioanalytical applications. An improved operational robustness would facilitate rapid and widespread adoption of liquid scanning probes in research as well as in diagnostics.

## I. INTRODUCTION

AKIN to scanning probes, such as atomic force microscopes (AFM), and scanning tunneling microscopes (STM) used for applications in e.g. metrology, liquid scanning probes enabling localization of liquids on biological samples, are poised to be increasingly used in biomedical research and medical diagnostics. [1], [2]. Liquid scanning probes enable local interaction with standard biological substrates, such as Petri dishes, microtiter plates and microscope glass slides. By scanning the probe across a substrate, distinct areas of interaction can be selectively chosen and interrogated.

Different variants of liquid scanning probes have been developed to localize liquids on specific regions on surfaces without cross-contamination between neighboring areas of interaction. Demonstrated implementations e.g. are based on the delivery of reagents in a second, aqueous phase to reduce mixing [3], or the release of minute amounts of reagents [4], [5]. Another approach is to create flow patterns in the gap between the liquid scanning probe and the sample by injection and simultaneous re-aspiration, to confine the flow of a reagent within a so-called hydrodynamic flow confinement (HFC). Associated methods impose no specific constraints on the properties of the applied liquids and are commensurate with the length-scales of standard substrate formats, as they enable interaction with areas ranging from tens of μm$^2$ to several cm$^2$ [6]. This principle is applied in e.g. the microfluidic fountain pen, the multifunctional pipette and the microfluidic probe (MFP) [7]–[11].

The MFP allows to establish a flow confinement of a liquid reagent (to as low as a few pLs) on a surface with a footprint at the scale of 100 μm × 100 μm (see Fig. 1B). During operation, a probe head with a flat apex is positioned at a distance of about 10-100 μm from a sample. From apertures at the center of the apex, a processing liquid can be injected and re-aspirated together with some surrounding immersion buffer, resulting in an HFC of the processing liquid on the sample surface. The use of the MFP and its capability for spatial and temporal multiplexing of liquids on surfaces has been demonstrated in the form of local multiplexed immunohistochemistry on tissue sections, local lysis of adherent cells and the creation of protein microarrays [1], [2], [12], [13], for example.

A key aspect limiting the transition of these liquid scanning probes to the life-sciences, biology and medicine is their lack of operational robustness. This lack of robustness is mainly caused by the slow or delayed detection of the occurrence of a failure by the peripherical operating instrumentation and monitoring units (see Fig. 1A). This can result in significant damage to the applied probe and the sample being probed. In this paper, with the exemplary example of the MFP, we identify two main operational failure modes of liquid scanning probes that rely on simultaneous injection and aspiration of liquids. As re-aspiration of a processing liquid is required in several scenarios, such failure modes can occur in scanning electrochemical microscopy (SECM) push-pull probes [14] and AFM-based probes used with liquids [15]. The strategies outlined here are hence broadly applicable to a range of liquid scanning probes. We here propose a design element, a microfluidic bypass channel, which can be operated in two configurations to prevent these failures by either enabling a passive response of the probe, or by extending capabilities of the monitoring unit to allow for rapid feedback.

Submitted for review on August 4, 2018. This work was supported, in part, by the European Research Council grant to G.V.K. (Project No. 311122, "BioProbe").

D. T. and G.V.K. are both at IBM Research – Zurich, 8803 Rüschlikon, Switzerland (email: gov@zurich.ibm.com). D. T. is additionally a doctoral student at École Polytechnique Fédérale de Lausanne (EPFL), Switzerland.

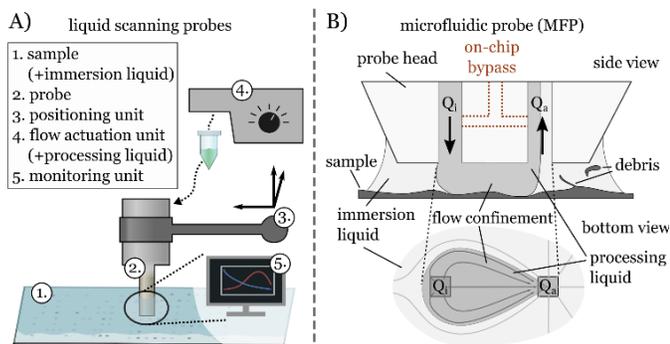

**Fig. 1.** Liquid scanning probes enable interfacing with immersed biological samples. A) Liquid scanning probes functionally comprise five components. B) The probe head of a microfluidic probe has a flat apex with at least two apertures. The probe head operates in proximity (~20 µm) to the surface of an immersed sample. By injection of liquid from the first aperture and re-aspiration at a higher flow rate from the second aperture, a reagent can be confined hydrodynamically to a specific region of the sample.

## II. FAILURE MODES OF LIQUID SCANNING PROBES AND STRATEGIES TO AVOID THEM

In liquid scanning probes, the fluidic structures and the instrumentation required to operate the probes are fairly simple, thus failures during operation largely result from the properties of the processing liquids and samples (e.g. agglomeration of solutes) and from improper handling (e.g. a misaligned placement of a sample). As biological samples and liquids can contain particles and bubbles and user-interactions are naturally prone to variation, addressing the problem of robustness is key to enable a wide-spread use of liquid scanning probes. We identified two main types of failure modes: a) the obstruction of one or more apertures resulting in a leakage of the processing liquid or damage to the components of the setup, and, b) a deviation from the desired gap distance resulting in an inhomogeneous interaction with the sample. As a means to mitigate the impact of these two main failure modes, we introduce the concept of a fluidic bypass structure between the injection and the aspiration channel. We make use of the analogies between fluidic and electrical circuits to illustrate and analyze the behavior of the introduced concepts. In the following designs and their analysis we assume that the flow of liquids is actuated by the setting of a controlled pressure level e.g. in a reservoir connected to a microfluidic channel. We emphasize that the presented concepts are equally applicable when imposing the flow rate with e.g. a syringe pump.

### A. Obstruction of apertures

Partial or complete obstruction of one or several apertures can be caused by either local elevations in the sample topography, or by particles and bubbles contained in the processing or the immersion liquid. As particles and small bubbles tend to follow the flow towards the aspiration aperture, this aspiration is likely to be clogged. During operation of the MFP and other liquid scanning probes that rely on re-aspiration of a processing liquid, an obstruction of the flow path between the injection and the aspiration aperture reduces the effective re-aspiration of the injected processing liquid. This can easily result in a leakage of the processing liquid into the surrounding immersion buffer and a contamination of regions of the sample, which should not have been in contact with the processing liquid. Fig. 2A illustrates this for partial obstruction of either the injection or the aspiration aperture. This failure mode is especially problematic in using a device with several parallel flow confinements [6]. Further, the obstruction of an aperture may lead to the build-up of increased pressure levels within the fluidic system and result in damage to hardware components, typically capillary-chip-interfaces, as well in leakage of the processing liquid in affected parts of the instrument. In the case where the aspiration aperture is used for picking objects from a surface, such as cells, from a surface, the build-up of high differential pressure levels might result in damaging those objects [15]. To the best of our knowledge, no strategies have been demonstrated to alleviate problems related to obstruction of channels.

Here we present a fully passive solution based on a liquid filled bypass channel between the injection and the aspiration channel in the probe head (Fig. 2B). An additional channel, the compensation channel, is connected to the bypass channel. Buffer is injected through the compensation channel at constant pressure to equalize the pressure on both sides of the flow-path element $R_4$ when the probe is positioned at the desired gap distance. If an obstruction of one or both apertures occurs, the differential pressure across the bypass increases, the pressure across element $R_4$ is not balanced and the flow of injected reagent is redirected through the bypass.

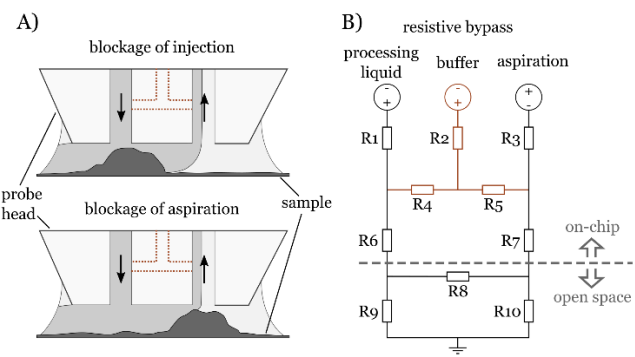

**Fig. 2.** Failure mode resulting from sample topography, or particles contained in the processing or immersion liquid and the proposed solution based on a resistive bypass. A) A blockage of either the injection or the aspiration aperture might result in a leakage of the processing liquid into the immersion liquid. B) Equivalent electrical model of a probe head with resistive bypass to mitigate failure due to obstruction of apertures.

### B. Variations of the gap distance between probe and sample

The second failure mode is induced by a deviation from the desired gap distance between the probe and the sample. Such a deviation is caused by scanning of the probe across a sample with variable topography, or by a mechanical misalignment between the scanning-plane of the probe and the surface of the sample. For given injection and aspiration flow rates, the interaction with the sample surface is specific for a gap distance [7], [16]. For small gap distances (<5 µm for the MFP), the probe and the sample might crash, while for large gap distances (>30 µm for the MFP, depends on the design of the probe and the ratio of aspiration/injection flow), the injected processing liquid might not be in contact with the sample surface (see Fig. 3A). In order for the HFC to interact with a sample in a repeatable manner, as required for e.g. assays with optical readout, working at a known and constant gap distance is crucial.

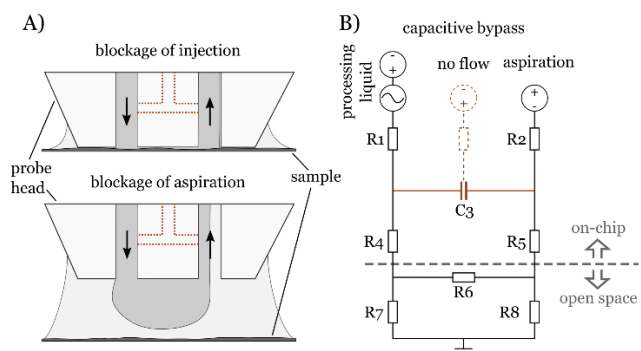

**Fig. 3.** Failure mode related to sample topography or a misaligned placement of a sample and proposed solution based on a capacitive bypass. A) In the case of a low gap distance, the probe might scratch and damage the sample. In the case of larger gap distances, the processed areas at the surface are not well defined or the processing liquid might not be in contact with the sample B) Equivalent electrical model of a probe head with capacitive bypass to mitigate failure due to obstruction of apertures.

AFM-based liquid scanning probe and probes related to scanning ion conductance microscopy (SICM) or scanning electrochemical microscopy (SECM) are readily suited for monitoring the gap distance by assessment of the deflection of a cantilever in the first case, or by measurement of a current between electrodes in the case of the latter ones. In contrast, for purely microfluidic liquid scanning probes, continuous monitoring of the gap distance poses a challenge, as many available approaches have limited compatibility with bioanalytical applications: optical measurement techniques, for instance, depend on the optical properties of the sample and are therefore not compatible with many assay. A measurement of electrical current between electrodes, as employed in SECM, imposes conditions on the applied buffer systems and requires the implementation of electrodes, which also limits the types of compatible applications [17]–[19]. A measurement of the hydraulic resistance of the gap between the probe and the sample requires a relatively high flow rate (above 100 µl/min) to obtain a quantifiable signal [20]. The sample surface thus is exposed to significant shear stress, which is incompatible with e.g. assays on surface-adherent cells. Other approaches are not commensurate with the range of gap distances of 5 µm to 100 µm, at which purely microfluidic liquid scanning probes are typically operated: AFM-related probes usually are operated at gap distances smaller than 1 µm [4], while e.g. a measurement of distances in a liquid environment using ultrasound waves has a limit of resolution at the scale of 30 µm [21].

We here present a method of measuring the gap distance between a liquid scanning probe and a sample based on the above introduced bypass channel, which provides a suitable range of measurement up to 25 µm and is well compatible with a wide range of bioanalytical assays. For this, we introduce a bubble of compressible gas into the bypass channel (see Fig. 3B) to change its behavior from purely resistive to capacitive. To stably hold the bubble in place, the geometry of the bypass channel is adapted to form a so-called coupling cavity. The pressure inside this coupling cavity can be adjusted via the compensation channel.

If the injection of reagent through the injection channel is now modulated at a certain frequency, the transduction of this modulation to the aspiration channel depends on the fluidic impedance of the fluidic network formed by the channels in the probe and the gap between the probe and the sample. A change in gap distance thus induces a change in phase shift of the transduced pressure signal. This phase shift is measurable with standard and simple imaging equipment, such as cell-phone cameras.

### III. THEORY

We designed and analyzed the bypass structures by drawing analogies between electric and fluidic circuits and making use of principles of voltage distribution across networks of electrical elements and the superposition of linear responses [22], [23].

#### A. *Resistive bypass – dc mode*

To study behavior of a resistive bypass, we performed a steady state nodal voltage analysis of the corresponding fluidic network (see Fig. 4A). We assume that flow is driven by three ideal pressure sources connected to the injection, the aspiration and the compensation channel, respectively. Following the Norton-theorem [24], all three ideal pressure sources are converted to ideal flow sources. For the combination of flow-path elements characterized in Fig. 4D, representing a standard probe design with resistive bypass, we numerically solved the system of linear equations to find the relative pressures at the nodes $\varphi_1 - \varphi_5$. With knowledge of the pressure drop across each of the involved flow-path elements, the actual flow through each element can be computed (see supplementary information section I for details on the calculation of hydraulic resistances). As shown in Fig. 4C, during normal operation, there is no flow across the first section of the bypass ($R_4$) and all the processing liquid flows through the injection aperture into the HFC and is then re-aspirated. There is a small inflow across $R_9$, as immersion buffer is drawn into the gap from either side, ensuring confinement of all the injected flow of processing liquid. To assess the response to obstruction of either one of the two apertures, the resistance of the flow-path element linked to either the injection aperture ($R_6$) or the aspiration aperture ($R_7$) was increased across two orders of magnitude. In response to an obstruction of the injection aperture, the nodal potential $\varphi_1$ increases and $\varphi_1$ and $\varphi_2$ are no longer equal. Consequently, a fraction of the injected processing liquid flows through the bypass, thereby reducing the effective injection flow of processing liquid into the gap. The effective injection flow of processing liquid further decreases due to the higher resistance of the obstructed injection flow path. The more severe the obstruction of the injection aperture, the more processing liquid passes through the bypass and the less flow enters the gap. In a scenario without a bypass channel, only the second effect, the reduction of the injection flow due to an increased resistance, occurs.

In the case of an obstruction of the aspiration aperture, in a probe with a resistive bypass, the nodal potentials $\varphi_2$ and $\varphi_3$ drop and again a share of the injected processing liquid flows through the bypass. As the negative relative pressure applied to the aspiration channel in absolute numbers is higher than the positive relative pressure applied to the injection channel, the pressure drop building up across the bypass channel in the case of an obstructed aspiration aperture can be higher, accordingly. If the value of $R_7$ exceeds a certain threshold, all injected processing liquid flows through the bypass channel and flow across $R_6$ is inverted, as immersion buffer is additionally drawn into the injection aperture (obstruction factor of 30 in the discussed design, see Fig. 4B).

In Fig. 4B, the flow across $R_6$ in the case of an obstruction of either the injection (faint colors) or the aspiration aperture (intense colors), normalized by the undisturbed flow across $R_6$, is plotted versus a factor of obstruction by which either $R_6$ or $R_7$ are multiplied. If there is an obstruction of the injection channel, in a probe with a resistive bypass, the reduction of the flow across $R_6$ is over 2-fold more pronounced than in a design without a bypass.

The benefits of a resistive bypass becomes more apparent for the case of an obstructed aspiration aperture: for a design without a bypass, the injection flow into the gap remains unaffected and processing liquid leaks into the surrounding immersion liquid. In a probe with resistive bypass, in contrast, leakage of the processing liquid is completely avoided.

### B. Capacitive bypass – ac mode

For probes with capacitive bypass elements, the network analysis facilitates the design and operation by highlighting the correlations between the applied relative pressure levels, the modulation frequency, the internal volume of the coupling cavity, the gap distance, and the resulting phase shift between the input modulation and the modulation transferred to the aspiration channel. Analysis of dc (constant flow) and ac (time-dependent flow) characteristics is performed separately and the results are superimposed (see Fig. 4E). Further, as the complex impedance is required to describe the frequency dependent properties of e.g. a capacitor (see supplementary information section I), all calculations are performed in complex space. There are two pressure sources relevant for the dc analysis: the first one supplies a constant positive relative pressure to the injection channel, whereas the second provides a constant negative relative pressure to the aspiration channel. In analogy to the above analysis of the resistive bypass, those two ideal pressure sources are converted to ideal flow sources. As there is no net current across the bypass ($Z_3$), this flow-path element is irrelevant for the dc analysis. Again, a system of linear equations is solved to find the dc components of the relative pressures at the nodes $\kappa_1 - \kappa_4$. For the ac analysis, the only source to be considered is a cos-type ac pressure source connected to the injection channel, the two dc sources are treated as short circuits. After conversion of this source to an ideal flow source and the solution of the corresponding system of linear equations, one obtains the relative amplitudes and phase shifts of the pressure modulation at the nodes $\kappa_1 - \kappa_4$.

This data, superimposed with the dc distribution of pressure across the network, allows us to compute the flow across any of the flow path-elements (assuming the system has equilibrated). On the basis of the described ac model, we analyzed how strongly a change in gap distance between probe and sample would change the phase difference between $\kappa_1$ and $\kappa_2$ and how well this effect would be observable. The pressure amplitude of the modulation generated by the ac source linked to the injection channel is transferred to the node $\kappa_2$ at a factor of about $10^{-2}$. Assuming input modulation amplitudes smaller than 100 mbar, a direct measurement of the transferred pressure signals using standard pressure sensors (resolution of ~0.25 mbar) would result in noisy signals. We thus chose to monitor the movement of the gas-liquid interface between the gas in the coupling cavity and the liquid in the aspiration channel, to which we refer as the rear interface (see Fig. 6C), to track the modulation of the pressure in the aspiration channel. The gas-liquid interfaces between the gas in the coupling cavity and the liquid in the injection and the aspiration channel reside in tapered structures, so that the radius of curvature of each interface is free to adapt to the corresponding differential pressure. Consequently, the acceleration of the rear interface is proportional to the pressure difference between the inner volume of the coupling cavity and the pressure at $\kappa_2$. We therefore assume that the movement of the rear interface exhibits the same relative phase shifts as the pressure at $\kappa_2$ itself. Based on the computed, time-dependent pressure levels at $\kappa_1$ and $\kappa_2$, as well as the geometrical boundary conditions and the properties of the involved fluids, we estimate the movement amplitude of the rear interface for a given combination of gap distance and modulation frequency (see supplementary information, section V). The surface plots in Fig. 4F display the estimated movement amplitudes for a range of modulation frequencies and gap distances, assuming a reasonable input modulation amplitude of 40 mbar. It can be seen that the movement amplitudes of the rear interface drop from over 100 µm to less than 20 µm, as the gap distance becomes smaller than 5 µm. This is due to the decrease of the relative pressure at $\kappa_2$ occurring when the probe is close to the surface of the sample, which results in an increased pressure drop across the rear interface. Under these conditions the rear interface adapts a relatively small radius of curvature and the changes induced by the modulation result in smaller relative changes of the interface position. Our model also predicts that the movement amplitudes get smaller for higher modulation frequencies. This occurs, because the analyzed network acts as a high-pass filter and more efficiently transfers the pressure modulation to the aspiration side at higher frequencies. A better transfer of the pressure modulation in turn results in a lower differential pressure across the coupling cavity and therefore in less movement of the gas-liquid interfaces. This effect shifts to higher frequencies as the volume of the coupling cavity is reduced. The color of the surface plots in Fig. 4F encodes the change in phase shift between the pressure levels at $\kappa_1$ and $\kappa_2$ for a change of the gap distance of 1 µm. Regions in which changes of the gap distance induce a high change of the phase shift are marked by more intense color. It is clearly observable that the changes in phase shift are more pronounced across a wider range of frequencies and gap distances for smaller coupling cavities. Movement amplitudes in the range of 100 µm can be resolved well with e.g. low-cost optical equipment (see subsequent section). Our model therefore suggests that a simple optical tracking of the rear interface should allow an observation of relative changes of the phase shift induced by changes of the gap distance for gap distances above 5 µm, with higher movement amplitudes to be expected for low frequencies.

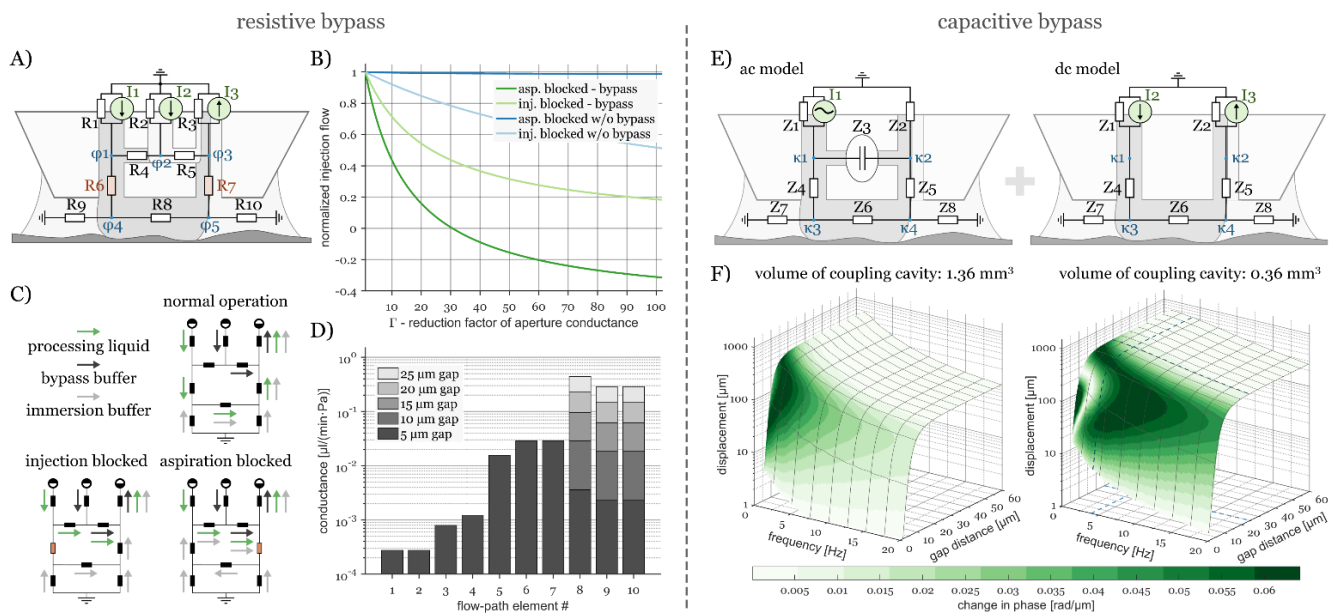

**Fig. 4.** Analysis of microfluidic probe heads with resistive or capacitive bypass structures. A) Equivalent electrical model of a microfluidic probe head with a resistive bypass. B) Plot of normalized flow through the injection aperture plotted over the factor of aperture-blockage for a device with resistive bypass (green) and a device without bypass (blue). The dynamic response of a probe head with resistive bypass to the event of an obstruction significantly reduces the risk of leakage of processing liquid. C) Flow path of liquids in a probe head with resistive bypass during normal operation and blockage of either the injection or the aspiration channel. D) Conductance of the individual flow-path elements of an exemplary probe head with resistive bypass. E) Equivalent electrical model of a microfluidic probe head with a capacitive bypass. The ac and dc behavior of the system was characterized separately and superimposed. F) Transfer of a sinusoidal pressure modulation from $\kappa_1$ to $\kappa_2$ in two designs featuring coupling cavities of different volumes.

## IV. EXPERIMENTAL DETAILS

### A. Probe heads and experimental platform

The MFP platform consists of five basic components: a sample holder with a sample covered by immersion liquid, a probe head, an XYZ positioning system to position the probe accurately relative to the sample, an imaging system, and a liquid flow actuation system.

The probe can be made from various materials to be suited for a specific application. In the discussed case the probe is a silicon glass device [12]. Channel structures were photo-lithographically defined and then etched into the silicon at a depth of 100 µm using deep reactive-ion etching (DRIE). Subsequently, the silicon wafer was anodically bonded to a BF33 glass wafer (1.3 kV, 475 °C), to seal the channel structures and single probes were then obtained by dicing. After dicing, the apex of the probe was lapped and polished on a wafer polishing tool (LP50, Logitech, UK) to obtain a well-defined and smooth surface.

The positioning system consists of three linear stages (Zaber Technologies Inc., Canada) with a positioning resolution of 0.05 µm. The sample can be scanned in the XY-plane relative to the probe, while the probe itself moves only along the Z-axis, in order to stay aligned with the optical axis of the microscope. Imaging of the HFC is performed from underneath, through a transparent sample, using an inverted microscope (here: Lumascope LS620, Etaluma Inc., USA). A dye, e.g. fluorescein, can be added to the processing liquid to facilitate the visualization of the HFC.

The tubing (1/16 FEP tubing, IDEX H&S, USA) is connected to the probe by a circular microfluidic connector (Dolomite Microfluidics, UK). Each fluidic channel is connected to a fluidic reservoir, in which the relative pressure is controlled using a pressure control system (EZ-Flow, Fluigent SA, France).

Only DI water was used as liquid in the described experiments. For better contrast, food colorant and fluorescein were added, as needed. We used a microscope glass slide as a dummy-sample. The glass slide had 250 µm thick slabs of PDMS on it, which allowed us to selectively obstruct either the injection or the aspiration aperture.

### B. Experimental setup for evaluating the capacitive bypass

In order to modulate the relative pressure in the reservoir connected to the injection channel in the experiments on the capacitive bypass, the reservoir was connected to a 2/3-way switch valve (24 V solenoid valve, Neptune Research & Development Inc., USA). The valve linked the reservoir to two gas-filled buffer containers with a volume of 0.5 liters each, one supplying positive relative pressure and the other one negative relative pressure (see supplementary information section IV for details). The pressure in the buffers was controlled by a pressure control system (EZ-Flow, Fluigent SA, France). The buffers were required to prevent an interference between the alternating relative pressure in the reservoir linked to the injection channel and the pressure control system, which is configured to maintain a constant pressure at its output ports.

The 2/3-way valve was supplied with power from a laboratory power supply and connected to ground across a bipolar transistor, which was switched by a square-wave signal generated with a waveform generator (33511B, Keysight Technologies, USA).

To record videos of the coupling cavity at a frame rate of 240 fps during operation of the MFP, we used a Galaxy S7 cellphone (Samsung Electronics, South Korea) with a 20× clip-on lens. Interface movements could be reliably tracked down to movement amplitudes of about 80 μm.

### C. Analytical modelling

The described systems of linear equations were solved using Matlab (MathWorks, Inc., USA). Also, the interior-point optimization routines used to find suitable input pressure levels, as described in supplementary information section III, were setup and run in Matlab.

### D. Image processing and measurement of the phase shift

To assess the phase shift between $\kappa_1$ and $\kappa_2$ in probes with a capacitive bypass, we analyzed the recorded videos of the coupling cavity to track the movement of the front interface (same relative phase shift as $\kappa_1$) and the rear interface (same relative phase shift as $\kappa_2$). The optical flow between two subsequent frames of the video was computed using Matlab, which allowed us to extract relative velocity amplitudes in defined regions of the videos [25]. This was performed for both, the front and the rear interface, then the acquired data was transformed to spectral space and finally the phase information at the modulation frequency was extracted for each interface from the corresponding fast Fourier transform (FFT) bin.

## V. RESULTS AND DISCUSSION

### A. Resistive bypass for compensation of obstructions of apertures

To test and demonstrate the functionality of the resistive bypass structures, we added food colorant to the liquids used in the experiment. This allows the tracking of specific liquids within the fluidic network. The processing liquid was colored green, the immersion buffer blue and the bypass buffer, which is injected through the compensation channel, was left transparent. Fig. 5A shows a photograph of a probe with a resistive bypass. The conductance of the individual flow-path elements of this design is displayed in Fig. 4D. The major share of the overall resistance of the bypass is attributed to the first section of the bypass ($R_4$), in order to limit the flow of buffer from the compensation channel into the injection channel in situations when the distance between probe and sample is larger than the desired gap distance.

During normal operation (see Fig. 5B), the processing liquid (green) flows through the injection channel (left hand side in Fig. 5B) and is re-aspirated together with some of the surrounding immersion liquid (blue) through the aspiration channel (right hand side in Fig. 5B). There is no flow across the first section of the bypass channel ($R_4$) and only flow of buffer (transparent) in the second section of the bypass channel ($R_5$).

We obstructed the apertures one after another by approaching the probe head towards a slab of PDMS of 250 μm thickness, which we positioned on the microscope glass slide that served as the sample surface in the described experiments. By centering the apex of the probe on the edge of the slab of PDMS, one aperture could be blocked selectively while leaving the other one unaffected.

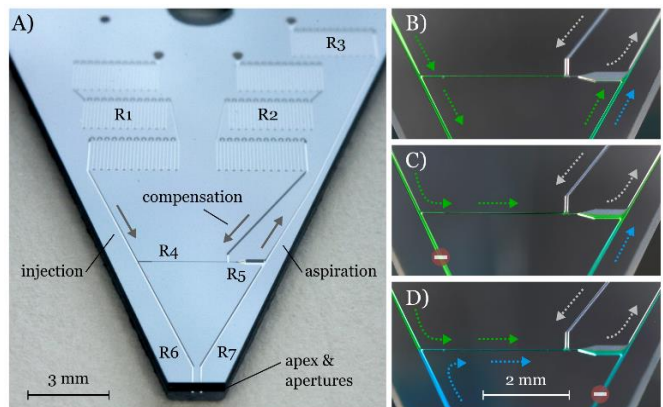

**Fig. 5.** Resistive bypass for dc configuration. A) Probe head with channels etched in silicon. B) Normal operation: no flow across first section of the bypass, only flow of bypass buffer through second half of bypass. C) Complete obstruction of the injection aperture: processing liquid (green) flows through the bypass. D) Complete obstruction of aspiration aperture: immersion liquid (blue) is pulled into the injection aperture and through the bypass together with the processing liquid.

If there is an obstruction of the injection aperture, flow in the injection channel starts stagnating and the relative pressure in the injection channel increases. Consequently, the total pressure drop across the bypass channel increases and the injected processing liquid is partially redirected through the bypass channel (see Fig. 5C). The probe can be brought in a full contact with the slab of PDMS and then be retracted subsequently without processing liquid leaking into the surrounding immersion liquid. In the case of a total obstruction, all injected processing liquid is redirected through the bypass.

When bringing the aspiration aperture gradually closer to the surface of the slab of PDMS, the aspiration flow across $R_7$ drops and the relative pressure in the aspiration channel gradually decreases. Again, this results in a higher total pressure drop across the bypass and a partial re-routing of the flow of injected processing liquid through the bypass channel. If the obstruction of the aspiration aperture is severe, all injected processing liquid is redirected through the bypass channel together with some immersion buffer, which enters the injection channel through the injection aperture (see Fig. 5C). Also the aspiration aperture can be brought into complete contact with the PDMS obstacle and then be retracted again without leakage of processing liquid into the immersion buffer.

### B. Capacitive bypass for continuous distance sensing

Fig. 6A depicts the probe used for demonstrating the concept of measuring the gap distance by observing the phase shift between the modulation of pressure in the injection and the aspiration channel. The coupling cavity has two tapered regions in order to let the front and the rear interface freely adapt to the differential pressure that drops across them [26]. The tapered region hosting the front interface is designed with a larger angle of inclination, as the differential pressure across the front interface is lower and this interface therefore adapts to larger radii of curvature than the rear interface. The drain channel and the probing channel did not have any function in the described experiments.

On the basis of the above considerations, we chose a probe design with a relatively small coupling cavity with a volume of 0.36 mm$^3$ and a modulation frequency of 4 Hz. This modulation frequency was found to be a good trade-off between resulting signal amplitudes, observable changes of the phase shift and the detectability of the signal with respect to noise from the environment. For tracing the phase shift between the pressure modulation in the injection channel and the pressure modulation in the aspiration channel, we decided to track the movement of the front and the rear interface using a video camera. By inferring the phase of the excitation signal and the transduced signal from the same video, we circumvent the problem of having to synchronize signals from multiple measurement devices.

We set the relative pressures in the reservoirs such that the mean injection flow rate was 3 µl/min and the mean aspiration flow rate 9 µl/min. We then recorded the movement of the front and rear interface for 5 s at a given gap distance, to integrate over 20 modulation periods. For gap distances below 25 µm, we randomly chose the order of the measurement points, to exclude hysteresis effects. The measured relative phase shifts between the front and the rear interface show a clear correlation with the gap distance. A change in phase shift is traceable up to gap distances of about 25 µm under the described experimental conditions. This covers the typical operating range during MFP experiments. In Fig. 4F we highlighted the chosen experimental conditions and the observed measurement range with dashed lines in the corresponding surface plot of the modulation transfer function. The observed measurement range is in good agreement with the theoretical prediction.

In order to fit the analytical model to the recorded experimental data, two modifications have to be made. The first modification is required, because the contact angle between the materials the probe is made of and water is smaller than 90°. As a consequence, liquid wets the inner walls of the coupling cavity, which gives rise to a bypass flow of liquid through the coupling cavity [27]. This has to be taken into account by adding a resistive element to $Z_3$ in parallel (see supplementary information, section VI). A second modification is required, as in the actual experiment there are parasitic capacitances spread across the fluidic network. We take these into account by adding a capacitive element in parallel to $Z_2$. By treating the properties of those additional elements as free fit parameters, the model can be fitted to the experimental data (see Fig 6B). To fit the experimental data, the value of the resistive element acting in parallel to the coupling cavity is estimated to be $1.25 \cdot Z_4$ and the sum of all parasitic capacitance in the network is estimated to be 1.23 times higher than the capacitance of the coupling cavity, corresponding to a volume of 0.44 µl. The values of these two fit parameters do not conflict with any aspect of the physical setup and are intuitively reasonable. For further refinement of the model, one might consider taking into account the dampening effects of viscous shear stress on the motion of the gas-liquid interfaces.

## VI. CONCLUSION AND OUTLOOK

We introduced a design element, a microfluidic bypass channel, which significantly improves the operational robustness of liquid scanning probes relying on re-aspiration of a processing liquid. The bypass channel can be operated in two configurations, each addressing one of the two main operational failure modes: in the dc mode, the bypass channel enables the fluidic network to intrinsically react to an obstruction of apertures and prevents leakage of the processing liquid into the immersion liquid. In the ac mode, the bypass enables continuous monitoring of the gap distance without the need for sophisticated sensing equipment and without adding constraints on the performed experiments. Even though the ideal geometry of the bypass channel is different for each mode, a bypass designed to be used in ac mode can be used just as well in dc mode.

While the performance of the bypass in dc mode fully meets the requirement of preventing leakage of processing liquid into the immersion liquid, if one or several apertures are obstructed, we envisage that monitoring of the gap distance in ac mode could be further improved. The resulting range of measurement from 5 µm to 25 µm could be enhanced by adding hydrophobic patches to the coupling cavity, to suppress its resistive behavior. Also, additional capacitive flow-path elements could be added to obtain e.g. local extrema of the phase transfer function at specific gap distances. Further, a higher resolution in tracking the interface positions by means of e.g. electrode arrays, or optical line sensors would increase the range of operation in terms of gap distance and modulation frequency.

We further envisage that the dc and ac functionality can be combined to directly enable leakage free operation at a constantly monitored gap distance: the differential pressure building up due to an obstruction could be employed to displace the gas in the coupling cavity and open up a resistive flow-path for internal bypassing of the processing liquid. The action of such a design element would be fully reversible and repeatable, as the gas volume would move back to its original position and close the bypass flow path for the processing liquid upon normalization of the differential pressure levels.

Bypass elements might also be used for rapid switching between different processing liquids [27]. In the context of more specific applications, the functionality of a fluidic bypass could be enhanced with additional functional elements, as e.g. spheres and bi-stable membranes. Such functional elements could open or close specific flow paths depending on the differential pressure dropping across the bypass. This would enable the probe to react to certain events with specific actions, without the need for any sensing or control infrastructure.

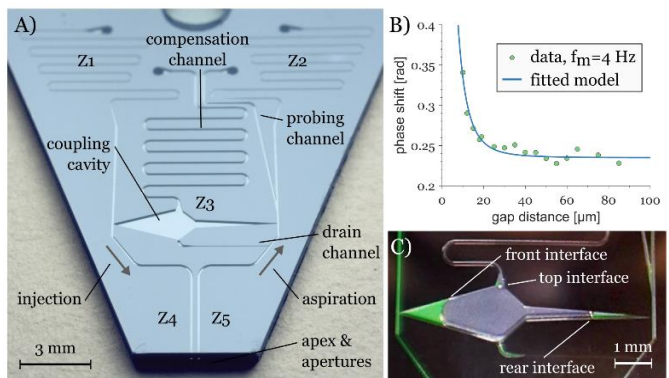

**Fig. 6.** Capacitive bypass for continuous monitoring of the gap distance. A) Probe head with channels and capacitive bypass etched in silicon. B) Differential phase shift between the movement of the front and the rear interface recorded at 4 Hz (dots) and fitted theoretical model (line). C) Air-filled coupling cavity during the experiment.

The presented concepts will help to improve the operational robustness, applicability and versatility of liquid scanning probes relying on the re-aspiration of liquid, while being compatible with a wide range of device concepts and bioanalytical applications. We hope the presented structures and methods help to pave the way for a reliable, long-term operation of such probes in diagnostic and analytical applications.

APPENDIX

- Supplementary information on analytical models, the experimental setups and corresponding methods.
- Video "Cavity_4Hz.mp4": recording of the coupling cavity (bypass in ac mode) at a gap distance of 20 μm and a modulation frequency of 4 Hz.

ACKNOWLEDGMENTS

We acknowledge financial support by the European Research Council (ERC) Starting Grant, under the 7th Framework Program (Project No. 311122, "BioProbe"). We thank Federico Paratore, Xander F. van Kooten, Robert D. Lovchik, Ute Drechsler, Marcel Bürge and Yuksel Temiz for stimulating discussions and support. Prof. Philippe Renaud (EPFL), Dr. Emmanuel Delamarche, and Dr. Walter Riess are acknowledged for their continuous support.

# Fluidic bypass structures for improving the robustness of liquid scanning probes

## Supplementary information to:

David P. Taylor and Govind V. Kaigala

### I. ESTIMATION OF PARAMETERS OF FLOW-PATH ELEMENTS

To calculate the hydrodynamic resistance of individual flow path elements, we used the formulas listed as follows:

- For elements with round cross-section (capillaries) [28]:

$$R_{round} = \frac{8 \cdot \mu \cdot l}{\pi \cdot r^4}$$

Here, $\mu = 1.002 \cdot 10^{-3}$ Pa·s is the dynamic viscosity of water at room temperature, $r$ is the cross-sectional radius of the flow-path element and $l$ is its length.

- For elements with rectangular cross-section (channels) [29]:

$$R_{rect} = \frac{4 \cdot \mu \cdot l}{w \cdot h^3} \cdot \frac{1}{\frac{1}{3} - \frac{64 \cdot h}{\pi^5 \cdot w} \cdot \tanh\frac{\pi \cdot w}{2 \cdot h}}$$

$w$ is the width and $h$ the height of the flow-path element under consideration, with $w > h$.

- To estimate the hydrodynamic resistance of flow-path elements in the gap between probe and sample, we assumed purely radial flow and one central injection emitting liquid at the flow rate $Q$ [28]. On the basis of continuity and assuming no-slip boundary conditions at the probe and the sample, it can be stated that

$$Q = \int_0^h dz \int_0^{2\pi} r d\theta A(r) z(h-z)$$

Here, the $z$-axis is oriented perpendicularly to the two parallel plates, $z = 0$ is at the surface of the sample and $A(r)$ is the area of the bounding surface at radius $r$. Integration over the gap distance h yields:

$$\vec{u}(r,z) = \frac{3Q}{\pi h^3} z(h-z) \vec{e}_r$$

Inserting this expression into the creeping flow approximation of the Navier-Stokes equations gives

$$\frac{\partial p}{\partial r} = -\mu \frac{\partial^2 \vec{u}}{\partial z^2}$$

and integration results in an expression for the pressure $p$ at a given radius $r$

$$p(r) = \frac{6\mu Q}{\pi h^3} \ln\left(\frac{r_1}{r_2}\right)$$

Therefore

$$R_{gap} = \frac{6\mu}{\pi h^3} \ln\left(\frac{r_1}{r_2}\right)$$

We assumed $r_1 = 100$ μm and $r_2 = 300$ μm for the HFC ($R_8$ and $Z_6$) and $r_1 = 100$ μm and $r_2 = 600$ μm for the elements linking to the edge of the apex ($R_9$, $R_{10}$ and $Z_7$, $Z_8$).

- To estimate the capacitance of the volume of air in the coupling cavity, we assume ideal gas laws and fully isothermal compression. The compressibility of the gas inside the cavity therefore is $k = -1/V \cdot \partial V/\partial t = 1 \cdot 10^{-5}$ Pa$^{-1}$. If the pressure applied through the liquid to the gas in the cavity changes, the volume of liquid inside the cavity changes:

$$Q = C \cdot \frac{d\Delta p}{dt}$$

$Q$ denotes the flow of liquid and $C$ the capacitance of the cavity. Insertion of $k$ yields:

$$Q = -\frac{dV}{dt} = -V \cdot \left(-\frac{1}{V}\frac{\partial V}{\partial p}\right) \cdot \frac{d\Delta p}{dt} = -kV \cdot \frac{d\Delta p}{dt}$$

It thus follows that

$$C = -kV$$

The impedance $Z_C$ of the capacitive element further depends on the modulation frequency $f_m$:

$$Z_C = \frac{1}{i \cdot 2\pi f_m \cdot C}$$

## II. SYSTEMS OF LINEAR EQUATIONS

Conversion of the ideal voltage sources to current sources (connected in parallel with the first flow-path element) following the Norton theorem [24]

$$I_1 = \frac{V_1}{R_1}$$

### A. Analysis of probes with resistive bypass

The system of linear equations for a probe with resistive bypass ($g_i$ denotes the conductance $1/R_i$):

$$\begin{pmatrix} g_1+g_4+g_6 & -g_4 & 0 & -g_6 & 0 \\ -g_4 & g_2+g_4+g_5 & -g_5 & 0 & 0 \\ 0 & -g_5 & g_3+g_5+g_7 & 0 & -g_7 \\ -g_6 & 0 & 0 & g_6+g_8+g_9 & -g_8 \\ 0 & 0 & -g_7 & -g_8 & g_7+g_8+g_{10} \end{pmatrix} \cdot \begin{pmatrix} \varphi_1 \\ \varphi_2 \\ \varphi_3 \\ \varphi_4 \\ \varphi_5 \end{pmatrix} = \begin{pmatrix} I_1 \\ I_2 \\ -I_3 \\ 0 \\ 0 \end{pmatrix}$$

### B. Analysis of probes with capacitive bypass

The system of linear equations for the ac analysis of a probe with capacitive bypass ($y_i$ denotes the admittance $1/Z_i$):

$$\begin{pmatrix} y_1+y_3+y_4 & -y_3 & -y_4 & 0 \\ -y_3 & y_2+y_3+y_5 & 0 & -y_5 \\ -y_4 & 0 & y_4+y_6+y_7 & -y_6 \\ 0 & -y_5 & -y_6 & y_5+y_6+y_8 \end{pmatrix} \cdot \begin{pmatrix} \kappa_1 \\ \kappa_2 \\ \kappa_3 \\ \kappa_4 \end{pmatrix} = \begin{pmatrix} I_1 \\ 0 \\ 0 \\ 0 \end{pmatrix}$$

The resulting values for the nodes $\kappa_1$ - $\kappa_4$ are phasors expressing the amplitude and phase of the respective node at the modulation frequency.

The system of linear equations for the dc analysis of a probe with capacitive bypass:

$$\begin{pmatrix} y_1 + y_4 & 0 & -y_4 & 0 \\ 0 & y_2 + y_5 & 0 & -y_5 \\ -y_4 & 0 & y_4 + y_6 + y_7 & -y_6 \\ 0 & -y_5 & -y_6 & y_5 + y_6 + y_8 \end{pmatrix} \cdot \begin{pmatrix} \kappa_1 \\ \kappa_2 \\ \kappa_3 \\ \kappa_4 \end{pmatrix} = \begin{pmatrix} I_2 \\ -I_3 \\ 0 \\ 0 \end{pmatrix}$$

### III. OPTIMIZATION ROUTINES FOR DEFINITION OF ADEQUATE INPUT PARAMETERS

*A. Determination of input parameters for a probe with resistive bypass*

We iteratively solved the system of linear equations in Matlab within an interior-point constrained optimization routine (fmincon) to find a workable combination of settings for all three sources with respect to the following conditions:
1. The optimization target was defined to be the minimization of the ratio of aspirated flow to the sum of the flows through the injection and the compensations channel. This makes the optimization settle in a local minimum in vicinity to the defined starting points ($p_1 = p_2 = -p_3 = 100$ mbar).
2. We postulate that the pressure $p_1$ of the source connected to the injection channel to be positive, i.e. flow is injected into this channel.
3. The flux $Q_7$ across flow-path element 7 should be three times larger than the flux $Q_6$ across flow-path element 6. This corresponds to an aspiration-to-injection-ratio of three in a classical probe without bypass. By experience this ratio results in a well-shaped HFC for common gap distances.
4. Further, the potential $\varphi_4$ should be negative relative to the ambient, to make sure there is inward flow to the gap between probe and sample and no processing liquid leaks to the immersion buffer.
5. The potentials $\varphi_1$ and $\varphi_2$ should be equal to suppress flow of processing liquid through the bypass during normal operation at the desired working distance.

At a working distance of 20 µm, for example, for the design characterized in Fig. 4D, suitable relative pressures would be $p_1 = 54$ mbar, $p_2 = 120$ mbar and $p_3 = -100$ mbar, resulting in total injection/aspiration flow rates of 1.5 µl/min through the injection channel, 3.5 µl/min through the compensation channel and 8 µl/min through the aspiration channel during normal operation.

*B. Determination of input parameters for a probe with capacitive bypass*

The above model of a probe with capacitive bypass is solved within an optimization routine, to find a combination of settings for the sources with respect to following conditions:
1. The optimization target was defined to be the minimization of the ratio of aspirated flow to the dc offset injection flow. This makes the optimization settle in a local minimum in vicinity to the defined starting points ($\hat{p}_1 = -p_3 = 100$ mbar).
2. We postulate that the pressure $p_1(t)$ of the source connected to the injection channel to be greater or equal to zero at all times, i.e. flow is injected into this channel.
3. The flux $Q_5$ across flow-path element 5 should be three times larger than the time averaged flux $\bar{Q}_4$ across flow-path element 4 to again reach an effective aspiration-to-injection-ratio of three.
4. Further, the time averaged potential $\bar{\kappa}_3(t)$ should be negative relative to the ambient, to make sure there is an overall inward flow to the gap between probe and sample and no processing liquid leaks to the immersion buffer.

For the proposed exemplary design (see Fig. Ap. 3C) for the admittance values of flow path elements of the specific design we analyzed) this analysis results in a constant offset pressure at the input of 40 mbar, which is superimposed by a sinusoidal modulation with an amplitude of 40 mbar, which results in a mean flow of 2.8 µl through the injection channel. At the aspiration a constant differential pressure of -99 mbar is applied, resulting in a mean flow of 9.8 µl/min.

S-12

## IV. MEASUREMENT OF RELATIVE PHASE SHIFTS

The measurement setup used for the assessment of the phase shift between the movement of the front and the rear interface in the coupling cavity is depicted in detail in Fig. Ap. 1. As mentioned in the main paper, buffers were required to avoid counter-productive controlling actions of the pressure control system as it would detects unstable pressure levels at its outputs if there were no buffers.

The flow sensors in each flow path are not needed in principle, but allowed to counter check the calculations of the resistance of the respective flow path and to monitor the system during the experiment more carefully.

The actual relative pressure levels set with the pressure control system during the experiment presented in the main paper in Fig 6B, were:

- Buffer 1:      410 mbar
- Buffer 2:      -210 mbar
- Aspiration reservoir:     -100 mbar (measured flow rate: 9.1 µl/min)
- Compensation reservoir:   ambient pressure

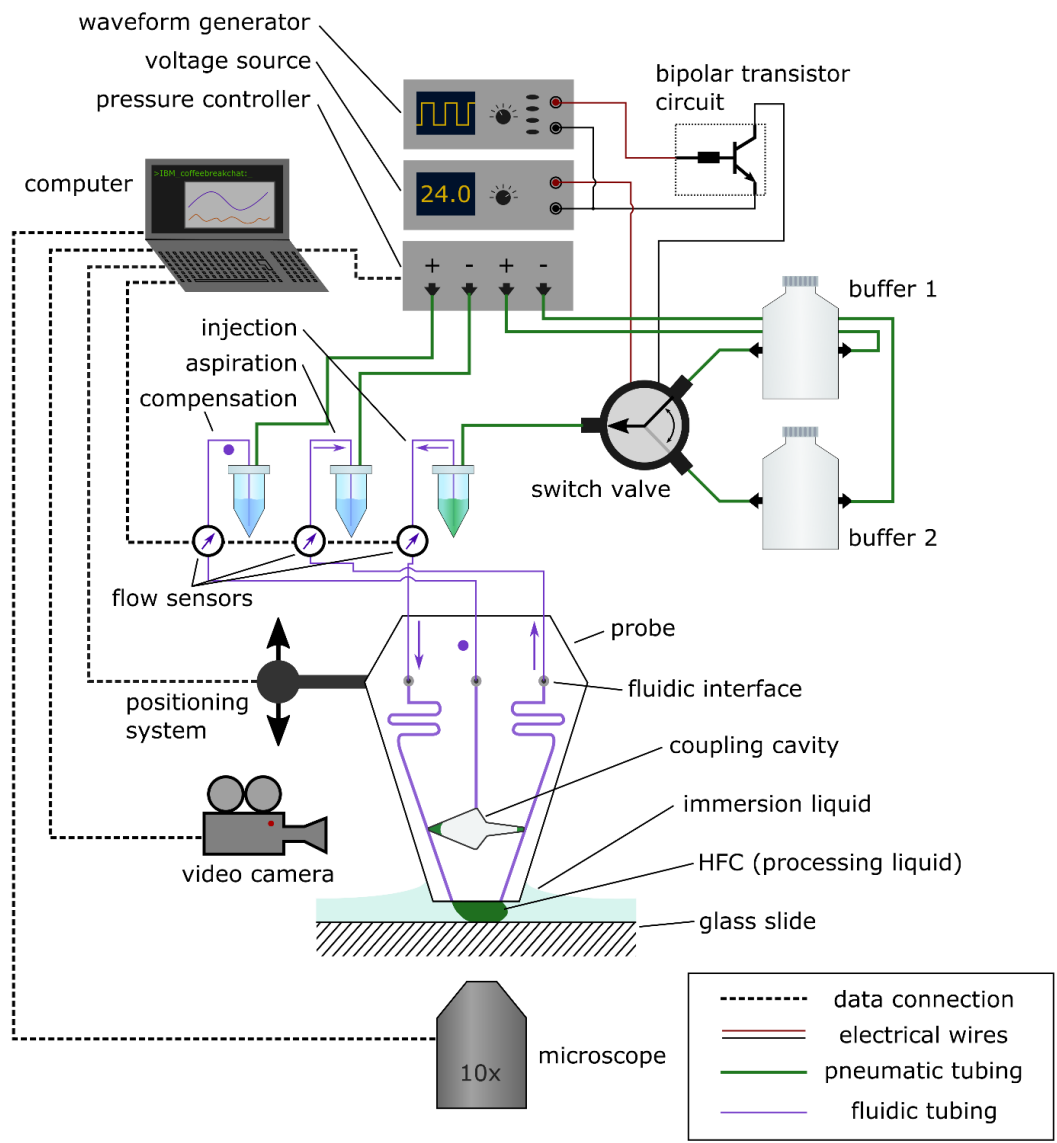

**Fig. Ap. 1.** Experimental setup for measuring the gap distance between probe and sample based on the movement of gas/liquid interfaces in the coupling cavity. In addition to the setup required for a probe with a resistive bypass, this setup includes instrumentation required to modulate the pressure in the reservoir connected to the injection channel and a video camera for recording a side view of the probe.

The negative relative pressure in buffer 2 allows the modulation amplitude to be increased. The relative pressures applied in the buffers greatly exceed the modulation amplitude of about 40 mbar suggested by the model. Nevertheless, as the model is in good agreement with the measured aspiration flow and a well-defined HFC was observed through the microscope (see video 1), we assume the actual pressure modulation in the injection reservoir to be in the range of the amplitude suggested by the model. We assume that only small fractions of the relative pressure levels present in the buffers get transferred to the reservoir, considering the dead volume and resistance of the tubing, the valve and the reservoir and the short switching times (250 ms, as the pressure is modulated at 4 Hz).

### V. CALCULATION OF THE RELATIVE MOVEMENT AMPLITUDE OF THE REAR INTERFACE

To estimate the movement amplitude of the rear interface we performed the following steps:
- We first estimated the inner pressure of the coupling cavity in a situation with constant injection and aspiration (no modulation) at a gap distance of 20 µm, by comparing the resulting positions of the front and rear interface with our observations from the experiment (see Fig 6C). Each interface adapts a position along the respective taper which results in a curvature of the interface corresponding to the pressure that drops across it. The opening width $d_0$ of the taper at the resting position of the interface is given by [26]:

$$d_0 = \frac{2 \cdot \cos(\theta - \psi)}{\frac{\Delta p}{\gamma} + \frac{h}{2 \cdot \cos(\theta)}}$$

Here, $\theta=45°$ is the estimated contact angle between the liquid and the walls of the coupling cavity, $\psi$ is the angle of inclination of the taper (15.5° for the front taper and 3.1° for the rear taper), h=100 µm is the etched depth of the coupling cavity and $\gamma = 75.75$ mN/m is the surface tension of the liquid.

As the reservoir connected to the compensation channel was mounted about 5cm above the surface of the immersion liquid surrounding the apex of the probe and no additional pressure was applied to this reservoir, we assume that a hydrostatic pressure of $p_c = 0.05 \text{ m} \cdot 0.1 \cdot 10^5 \text{ Pa/m} = 500$ Pa built up in the coupling cavity. With mean relative pressures of 128 Pa at $\kappa_1$ and of $-414$ Pa at $\kappa_2$, the front interface, according to the model, stabilizes at a width $d_{0f} = 340$ µm and the rear interface at a width $d_{0r} = 120$ µm. These interface positions are in good accordance with our experimental observations.

- The resting position of the rear interface is re-calculated for each examined gap distance by taking into consideration the calculated relative pressure at $\kappa_2$ and an internal pressure of the cavity of $p_c = 500$ Pa.

- We assume the response of the interface to be linear for small variations of pressure. To find the deviation of the interface from its resting position due to the modulation of the pressure, we evaluate the maximum change of pressure drop across the coupling cavity due to the modulation $dp = \max(p_{\kappa_1}(t) - p_{\kappa_2}(t))$. Then, using the above equation, the total change in position $dx$ is:

$$dx = \frac{d_0}{\Delta p} \cdot dp \cdot \frac{1}{\tan \psi}$$

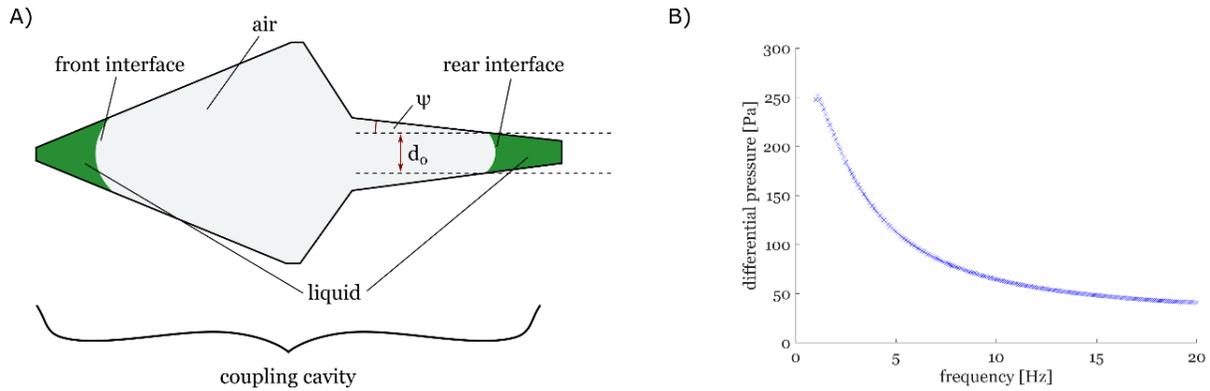

**Fig. Ap. 2.** Movement of the front and rear interface in the coupling cavity due to the pressure modulation. A) The opening width of the taper is lower for the rear interface as the pressure drop across this interface is higher and the resulting curvature of the interface is smaller. B) The differential pressure drop across the interface decreases with increasing modulation frequency, as the transfer of the modulation is more efficient at higher frequencies.

- Two effects impact $dx$ across the space of frequencies and gap distances: for small gap distances $\frac{d_0}{\Delta p}$ becomes very small (this derivative has a quadratic dependence of the pressure in the denominator) and thus the movement amplitudes get smaller, too. For higher frequencies, the modulation of the pressure is more efficiently transduced to the aspiration channel and the pressure difference across the cavity $dp$ becomes smaller (see Fig. Ap. 2B), which also in a reduction of $dx$.

## VI. MODIFICATION OF THE MODEL TO FIT THE EXPERIMENTAL DATA

When comparing the recorded experimental data with the analytical model presented in the paper, we found that a capacitive element has to be introduced in parallel to $Z_1$ or $Z_2$, in order to account for parasitic capacitances across the fluidic network and, more importantly, a resistive element connected in parallel to the capacitor of the coupling cavity has to be taken into account. This resistive element was present as well in the described experiments: as the liquid has a contact angle to the walls of the coupling cavity lower than 90°, a liquid film covers the walls of the coupling cavity and enables a direct flow of liquid between $\kappa_1$ and $\kappa_2$. As demonstrated in Fig. Ap. 3B, this resistive element leads to a significant change in the behavior of the fluidic network: the blue curve represents the behavior of the model fitted to the experimental data (with an additional capacitive element added to $Z_2$ and a resistive element added to $Z_3$) and the brown curve displays the behavior of the same model, but without the resistive element added to $Z_3$. The fitted model (blue curve) originally converges towards 0 for large gap distances and has to be shifted by an offset of 0.23 rad to fit the experimental data. The model without a bypass flow through the cavity (brown curve) approaches a constant value large gap distances. We therefore believe that the most accurate description of our experimental data would be a hybrid model, as the thickness of the liquid film at the walls of the cavity, and thus also the resistance of the bypass through the cavity, changes with the relative pressure that drops across the coupling cavity. We did not further investigate this, as the purpose of this study was to demonstrate the basic feasibility of assessing the probe-sample distance by observing relative phase shifts in the movement of the two interfaces. We expect that the total relative phase shift and thus the sensitivity and measurement range of the presented approach can be significantly improved by making regions of the coupling cavity hydrophobic to prevent a bypass flow of liquid.

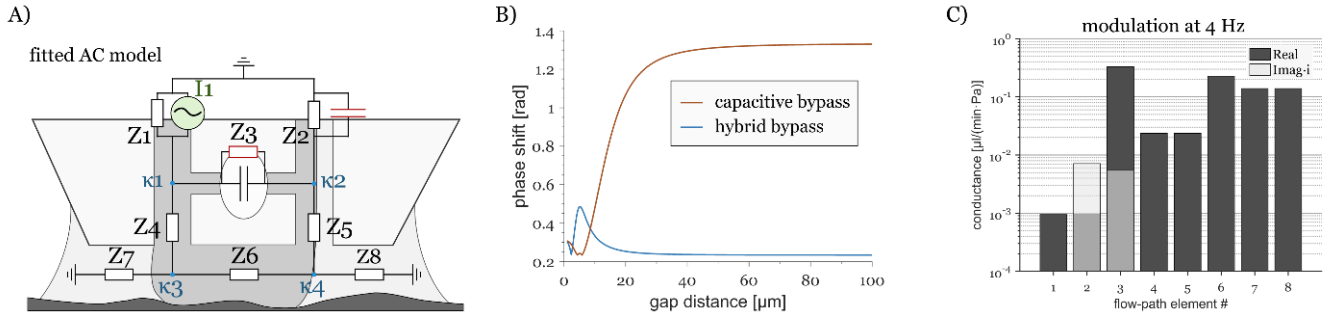

**Fig. Ap. 3.** Adjusted model for ac configuration. A) In order to fit the experimental data additional flow-path elements (red) were included in the model. B) Comparison of a model with parallel resistive element in the bypass (blue) and a model without parallel resistive element in the bypass (brown). The capacitor in parallel to $Z_2$ was kept for both models. C) Admittances of the flow-path elements with capacitive bypass (volume of coupling cavity is 0.36 mm$^3$). The admittance of elements with capacitive characteristics has an imaginary part (displayed in light grey color).